# Implementation and practical aspects of quantitative decision-making in clinical drug development

**Juan J. Abellan[1], Nicolas Bonnet[2], Alex Carlton[1], Paul Frewer[3], Heiko Götte[4], John-Philip Lawo[5], Jesper Madsen[6], Oliver Sailer[7], Guido Thömmes[8], Gaëlle Saint-Hilary[9,10], on behalf of the European Special Interest Group on Quantitative Decision-Making**

[1]GlaxoSmithKline, Brentford, Middlesex, UK, [2]Sanofi, Montpellier, France, [3]Oncology Biometrics, Oncology R&D, AstraZeneca, Cambridge, UK, [4]Merck Healthcare KGaA, Darmstadt, Germany, [5]CSL Behring, Marburg, Germany, [6]Novo Nordisk, Bagsvaerd, Denmark, [7]Boehringer Ingelheim Pharma, Biberach, Germany, [8]Grünenthal, Aachen, Germany, [9]Saryga, France, [10]Politecnico di Torino, Torino, Italy.

## 1. Introduction

For decades, biostatisticians have focused on statistical methods to make clinical trials and related analyses more reliable, to provide unbiased estimators of the treatment effect, to control the type I and type II errors of clinical study decisions, e.g. based on statistical significance. However, clinical drug development is becoming increasingly challenging, due to the high failure rate and the increasing costs (Arrowsmith et al, 2013). Antonijevic (2015) highlights new safety and efficacy requirements from the regulatory authorities, resulting in additional clinical trials, longer developments and increased costs. Thomas et al (2016) found that only 9.6% of the drug candidates in Phase I successfully reach Food and Drug Administration (FDA) approval between 2006 and 2015. Several factors were suggested to explain these results, such as unclear or heterogeneous population selection (Thomas et al, 2016), or portfolio prioritization decisions more focused on shortening the developments than increasing their probability of success (Antonijevic, 2015).

When conducting a clinical trial, there should be a clear link between the scientific question(s) of interest, the study design, the study results and the decisions made from these results, e.g. to continue or to stop the development. The definition of the development success could encompass different aspects, from the relevance of purely clinical characteristics of a drug (efficacy and safety) to its commercial attractiveness. While a substantial effort is put nowadays into the study design (inclusion/exclusion criteria, endpoints, assumptions for sample size…), less attention is paid to mapping results to decisions. For example, a statistically significant treatment effect does not support a decision to continue the development if this effect is lower than any clinically (or commercially) relevant value.

Also, each decision made during the clinical development must be carefully informed. While the amount and variety of information available to support these decisions increased, its complexity makes it challenging to synthesise it in a meaningful way (Welton, 2012).

Finally, since the true treatment effect will never be known, one must deal with the uncertainty about this true value given the data collected from the study (or development programme). The study design could be optimised to decrease this uncertainty, but will not be able to remove it. This is particularly true in early phase trials, where the sample size is restricted in order to both limit the number of participants exposed to and restrain the investments in drug candidates with a high likelihood of failure. Thus, the decision-making process must assess the risk of wrong decisions in the presence of uncertainty.

Quantitative decision-making (QDM) principles address these issues related to the mapping of results to decisions, the synthesis of information and the quantification of uncertainty. This leads to several topics for which biostatisticians have a critical role to play, and they are therefore ideally suited to drive this type of discussion.

Since the clinical drug development involves a succession of decisions to be made, QDM methods can be applied at various levels. At the study level, it can be used to properly design a study, and improve the GO/STOP decisions that are made either during the trial (for early termination or to adapt the design) or at its end. Establishing decision criteria ahead of the study is essential here to address the need for speedy decisions, potentially in real time. At the project level, QDM can be used to inform decisions to continue, adapt or stop a drug development programme based on results from previous studies. For instance, the predictive probability of success (PPoS) is a useful QDM tool for this purpose. At the portfolio level, QDM can be used to choose, prioritise and optimise the development portfolio, e.g. using the probability to reach market access or target sales within a predefined timeline.

This article is structured as follows: Sections 2 and 3 introduce the QDM frameworks. Their operating characteristics are presented in Section 4. Section 5 focuses on the communication with study teams, and finally a discussion and some concluding remarks are provided in Section 6.

## 2. Notation and a note on statistical inference

Here we introduce some notation while providing a brief overview of basic ideas of statistical inference. Let $\delta$ be the unknown quantity of interest. In clinical studies, $\delta$ could be the effect under a certain treatment or a difference between two treatments, with positive values of $\delta$ representing improvement. In the portfolio setting, $\delta$ can be the expected net present value (eNPV) (Parke, 2017).

We will make statistical inference from the data collected in our study to learn about $\delta$. If we use frequentist methods, then we will typically construct our likelihood function $L(\delta)$ and will obtain a point estimate $\hat{\delta} = \text{argmax}\, L(\delta)$, and a (say 95%) confidence interval $(\hat{\delta}_L, \hat{\delta}_U)$, the length of which will inform us about the uncertainty around $\delta$. If the inferential process happens within the Bayesian framework, we will have a prior distribution $p(\delta)$, which may be informative (e.g. if based on existing data), and it will be updated with the data from the study to produce the posterior $p(\delta \mid data)$. We can then produce relevant summaries of the posterior distribution, e.g. the mean as point estimate and a (95%) credibility interval, as well as probability statements of the form $\Pr(\delta > c \mid data)$ to ascertain our confidence about $\delta$ being larger than $c$.

In addition to statistical inference, in the frequentist paradigm it is also common to carry out hypothesis testing, for example to contrast $H_0: \delta = \delta_0$ vs. $H_1: \delta \neq \delta_0$. The resulting p-value of the test informs about the compatibility between data and the null hypothesis, with low p-values suggesting low compatibility. If the test is rejected at a significance level $\alpha$, then $p < \alpha$, or equivalently $\hat{\delta} > c_\alpha$, where $c_\alpha$ is some critical value on the point estimate scale.

## 3. QDM Frameworks

### 3.1. Constructing QDM frameworks

Without loss of generality, we introduce here simple quantitative decision-making (QDM) frameworks at the study level. Extensions to more complex settings and to other contexts are presented in the next Section.

A QDM framework is a set of criteria, established before the study starts, mapping study outcomes to decisions concerning next steps in the drug development process. Ultimately, the key decisions to be made upon study completion are of the type 'GO' and 'STOP', broadly indicating to progress or to terminate the development of the new drug, respectively. Positive evidence of efficacy/safety will typically map to a GO decision, whereas negative evidence will be mapped to a STOP decision. Quantitative criteria are used to establish what constitutes positive or negative evidence in favour of the new drug under development. Such QDM criteria are key to determine or optimise important aspects of the study design (e.g. sample size) to increase the chances of the study to deliver useful results for decision making. The assessment at the design stage of the *operating characteristics* of a QDM framework and how they can feed back the study design, will be discussed in detail in Section 4. In this Section we will focus on considerations to build a QDM framework.

In the past we have mapped results to decisions based on statistical significance, which can be seen as the following QDM framework based on p-values and a pre-specified level of significance $\alpha$:

- If $p < \alpha$, then GO
- If $p \geq \alpha$, then STOP

This familiar framework considers a statistically significant result of a hypothesis test ($p < \alpha$) as positive evidence that the drug is efficacious and maps it to a GO criterion. If the GO criterion is not met, then the decision is to STOP. While this framework may be appropriate for confirmatory studies, in earlier stages of development where we are still learning about the drug, the use of hypothesis testing and, as a consequence, of this framework may be more questionable. Since a p-value on its own says little around the size of the treatment effect and associated uncertainty (see Wasserstein and Lazar, 2016; Wasserstein *et al*, 2019), decision criteria based on p-values may increase the chance of erroneous decisions.

When developing a new drug, we want to build confidence that it has a clinically relevant impact. Available knowledge of the disease and the drug landscape in the indication of interest may be used to establish the efficacy (and safety) characteristics the new drug should have to add value for patients. We could use that information (often reflected in the target product profile (TPP) once available) to establish criteria for decision making. Let $MV$ be the minimum value of the treatment effect ($\delta$) stated in the TPP of the new drug. The stronger the evidence that $\delta > MV$, the higher confidence we want to make a GO decision. Based on this, we could consider a decision criterion mapping high confidence on $\delta > MV$ to a GO decision. But how do we assess our confidence on a sizeable treatment effect? It seems sensible to use inferential tools to quantify the uncertainty around the treatment effect $\delta$ being at least $MV$, such as suitable confidence/credibility intervals or posterior probability statements. With these elements we could build a QDM framework of the form:

- If we are at least X% confident that $\delta > MV$, then GO
- If we are at least Y% confident that $\delta < MV$, then STOP
- Otherwise, CONSIDER

This framework requires strong evidence of $\delta > MV$ to GO and strong evidence of $\delta < MV$ to STOP. There is also a CONSIDER decision outcome to cover the situation where both the GO and STOP criteria are not met. Study teams may further elaborate the QDM framework by explicitly stating what additional evidence will be considered for decision making if the study results end up in the CONSIDER zone, and how it will be mapped to one of the two decisions GO or STOP that ultimately need to be made.

Fisch et al (2014) propose a combination of the two QDM frameworks above, i.e. based on statistical significance and clinical relevance, for decision making in proof of concept studies:

- If we are at least X% confident that $\delta > MV$ and $p < \alpha$, then GO
- If we are less than X% confident that $\delta > MV$ and $p \geq \alpha$, then STOP
- Otherwise, CONSIDER

More elaborated QDM frameworks can be produced by considering additional relevant sizes of the treatment effect to construct decision criteria. Lalonde et al (2007) and Frewer et al (2016) consider, in addition to the $MV$ (which they refer to as *low reference value*), a target value $TV$ for the size of the treatment effect ($TV > MV$) for the new drug to have a more differentiated profile. The QDM framework proposed by Frewer et al (2016) map high confidence on $\delta > MV$ to a GO decision and high confidence on $\delta < TV$ to a STOP decision. Their QDM framework could be written as

- If we are at least X% confident that $\delta > MV$, then GO
- If we are at least Y% confident that $\delta < TV$, then STOP
- If both conditions are simultaneously met, then STOP.
- If both conditions are simultaneously not met, then CONSIDER

With two conditions for decision making (based on $\delta > MV$ and $\delta < TV$), one needs to think of decisions for when none of the two conditions is met and for when both are met, even if in practice there might be a small chance for the latter to happen (e.g. in small studies). Lalonde et al (2007) and Frewer et al (2016) mapped the study outcome of not meeting any of the two conditions to a CONSIDER decision, whereas simultaneously meeting both was mapped to a STOP. Other decision options could also be envisaged for this last case where we end up having high confidence on $MV < \delta < TV$. For example, the study team could map it to a GO decision if the team weighs the first condition ($\delta > MV$) over the second ($\delta < TV$). It could also be mapped to a CONSIDER decision, if the team thinks that additional considerations may have a role to play before deciding whether to GO or STOP (Quan et al 2019). However, having the same CONSIDER label for two very different study outcomes (meeting both conditions and not meeting any) may create confusion in study teams. In this situation, perhaps using another decision label when the two conditions are met (e.g. INTERMEDIATE), may improve communication within the study team and avoid potential misunderstandings.

The choice of relevant sizes of the treatment effect (e.g. MV and TV) to establish a QDM is context-dependent and may be informed by several sources of information. In very early stages, the understanding of the biology of the disease, the mode of action of the new drug, the existence of preclinical data, etc. may be used to set reference values for the quantity of interest $\delta$. In later development stages, additional elements such as clinical experience on treating patients, knowledge of the competitor landscape in the same indication, information from patient organizations, etc. may be used to determine those reference values. In any case, scientific knowledge and judgement are as always crucial to this end.

Selection of the levels of confidence $X\%$ and $Y\%$ to GO or STOP will depend on the risks the study team is prepared to take, which will again depend on the setting (e.g. an indication with high unmet need vs. a competitive one) and the investment the decision is gating (e.g. entering phase II vs. phase III). It is nevertheless important to understand the impact of specific choices on the operating characteristics, i.e. the ability of the study to deliver good decisions. This may involve exploring several scenarios for the size of the treatment effect and assessing the probabilities of correct and incorrect GO/STOP as well as the probability of CONSIDER. The challenge is therefore to find the right balance between those probabilities and a sensible sample size to generate the evidence needed to make a good decision within a reasonable timeframe. For example, criteria requesting 95% confidence that $\delta > MV$ to GO and 95% confidence that $\delta < TV$ to STOP will reduce the risks of wrong decisions but will increase the chance to end up in a CONSIDER zone. Increasing the sample size may reduce the CONSIDER probability, but at the expense of increasing the cost and length of the study. It is also important to explore the interplay between a proposed QDM framework and other aspects of the study design. This may help to optimise the design. For example, if the risk of a wrong decision at the end of the study is relatively large, we may consider introducing an interim analysis as a mitigation strategy. Or a study team considering increasing the sample size of a study may realise that such an increase is not worth because it barely improves the operating characteristics.

### 3.2. QDM frameworks in more complex settings

In the previous Section we have focused on QDM frameworks based on the observed treatment effect from one efficacy endpoint. However, they could also be based on more than one efficacy endpoint, or on a combination of efficacy and safety endpoints (e.g. benefit-risk). In this case, the QDM frameworks presented above could be applied to a utility score that summarizes the benefit-risk balance in a single measure (Saint-Hilary et al, 2018). Alternatively, it could be expanded to a decision-tree where a GO, STOP or CONSIDER decision is associated to each combination of endpoint outcomes (Abellan et al, 2019).

At the program level, the design of phase II trials may be based on success probabilities for future phase III trials (Götte et al, 2015). Also, GO/STOP decisions may be made with a surrogate or short-term endpoint in early phase, while eventually the new drug will be assessed on clinical endpoints of interest in late phase (Hong et al, 2012; Sabin et al, 2014; Saint-Hilary et al, 2019; Wang et al, 2013; Götte et al, 2020).

Other decisions than GO/STOP decisions may also be of interest, for example selecting a target population or a dose (Dunyak et al, 2018; Rufibach et al, 2016; Götte et al, 2017). Decision criteria involving non-scientific or commercial aspects like cost, revenue, time or competitor information may be included too (Antonijevic, 2015). This is particularly true when decisions are made at the portfolio level, where the objective is to choose, among different strategies on the pipeline, the one that maximises the financial value and/or minimises the risks given some budget constraints (Kirchner et al, 2016; Patel et al, 2013; Thunecke and Elze, 2014, Graham et al, 2020). QDM frameworks could be built in such situations as well, using the same principles demonstrated above: thresholds are defined on quantitative criteria (e.g. Net Present Value, number of marketing authorisations) and decisions are made given the level of confidence that such thresholds are exceeded.

The examples above are not an exhaustive list and there are other settings that may also benefit from pre-specified QDM frameworks.

### 3.3. Examples

Let us consider examples, inspired by a real case-study, on the development of a drug in patients with mild to moderate Alzheimer's disease. The primary endpoint is the 11-item ADAS-Cog total score change from baseline after 24 weeks of treatment, assumed normally distributed. In the TPP, it is pre-specified that a minimal value $MV = 2$ points in the difference of means on the primary endpoint is clinically relevant to add value for the patient population, and a target value $TV = 3$ points corresponds to a substantial improvement compared to the standard of care. We suppose that the future development plan consists in one Phase II study followed by one Phase III study.

To illustrate the ideas presented in the previous section we consider two possible QDM frameworks that one could consider in this setting. In both examples, the QDM framework is assumed to be specified at the planning stage of the Phase II study to make the decision at the end of Phase II to pursue the development in Phase III (GO) or not (STOP).

**Example (a). QDM framework based on available Ph II evidence and uncertainties**

This framework is based on the *posterior probability* that the treatment effect is large enough, considering the available evidence and uncertainties (Phase II data):

- If $\Pr(\delta > MV \mid data\ in\ Ph\ II) > 70\%$, then GO;
- If $\Pr(\delta < TV \mid data\ in\ Ph\ II) > 90\%$ then STOP;
- If both conditions are simultaneously met, then STOP;
- If both conditions are simultaneously not met, then CONSIDER.

Hence, the decisions are made here based on the level of confidence that the new treatment adds value for the patient population. The calculation of the posterior probabilities above is based on a non-informative prior $p(\delta)$.

**Example (b). QDM framework based on available Ph II *and future* Ph III evidence and uncertainties**

This framework takes into account the *predictive probability* (or *assurance* (O'Hagan et al, 2005)) that the treatment effect will be large enough and statistically significant in Phase III. Therefore, it takes into account evidence and uncertainties from both past (Phase II) and future (Phase III) data.

As a first step, we determine the decision criteria that will be used at the end of Phase III to continue with a marketing authorisation application (GO after Phase III) or not. Similar to Fisch et al (2014), the GO criterion is based on statistical significance and clinical relevance considering the data collected in Phase III. Let $\hat{\delta}$ be the estimate of the treatment effect $\delta$ in the Phase III study. In this study, the null hypothesis $H_0$: $\delta \leq 0$ will be rejected at level $\alpha$ if $p < \alpha$, or equivalently $\hat{\delta} > c_\alpha$, for $c_\alpha > 0$ some critical value. We consider the following framework:

- If $\hat{\delta} > c_\alpha$ and $\Pr(\delta > MV \mid data\ in\ Ph\ III) > 80\%$, then GO after Phase III;
- Otherwise STOP after Phase III.

The second step is to set-up the criteria to decide at the end of Phase II to pursue the development in Phase III (GO) or not, based on the predicted chance of success in Phase III:

- If $\Pr(G0\ in\ Ph3 \mid data\ in\ Ph\ II) > 70\%$, then GO;
- If $\Pr(G0\ in\ Ph3 \mid data\ in\ Ph\ II) < 30\%$, then STOP;
- Otherwise, CONSIDER.

The decisions are made here on the level of confidence that the new treatment adds value for the patient population, but also that the development plan permits to show it. In other words, at the end of Ph II decisions will be made upon the uncertainty surrounding predicted Ph III results. Although this QDM framework has the advantage of considering all possible uncertainties, it requires that the Phase III study is already designed at the planning stage of the Phase II study, when the QDM framework is pre-specified. This is not always the case and may be difficult to achieve in practice, but in some cases even a rough idea of the sample size in Ph III may suffice.

## 4. Operating characteristics

The evaluation of operating characteristics in the planning phase is crucial to determine the usefulness of a QDM framework.

### 4.1. Conditional operating characteristics

*Conditional operating characteristics* allow assessing how often the decision rules of a QDM framework lead to good or bad decisions when the true treatment effect is assumed to be a fixed known value. They consist in calculating the probabilities of GO, STOP and CONSIDER decisions over a range of possible scenarios. Of particular interest are the probabilities of Correct GO and Incorrect STOP decisions, computed when assuming the true treatment effect is clinically meaningful; and the probabilities of Incorrect GO and Correct STOP decisions, computed when assuming there is no or a non-meaningful treatment effect (Table 1). It should be noted that the probabilities of Incorrect GO and Incorrect STOP decisions correspond to the false positive and false negative error rates, and are analogous to the type 1 error and type 2 errors when testing a null hypothesis in a frequentist framework (Walley et al, 2015; Chuang-Stein and Kirby, 2017). One error rate increases while the other decreases when changing the decision boundary, so both cannot be lowered simultaneously without sample size increase.

Conditional operating characteristics also permit to check, as recommended by Frewer et al (2016), that the probability of a CONSIDER decision is limited (for example, that it does not exceed 30%), otherwise the chance of not having a definitive result at the end of the study may be too high and the benefit of the decision-making framework may be lost.

|  |  | Scenario | |
|---|---|---|---|
|  |  | **Clinically meaningful treatment effect** | **Non-meaningful treatment effect** |
| **Decision** | **GO** | Correct GO | Incorrect GO |
|  | **STOP** | Incorrect STOP | Correct STOP |

**Table 1. Correct and incorrect decisions over possible scenarios**

If the analysis prior for $\delta$ is non-informative, some rates are fixed by the QDM framework. For example, in the example (a) presented in Section 3.3, the probability of STOP given the true treatment effect equals TV (Incorrect STOP) is fixed at 10%. The acceptable error rates can be set at a company-wide level, either fixed for all studies or varying according to criteria specific to the development program. For example, high unmet medical need could lead to liberal GO boundaries (high acceptable Incorrect GO rates and low acceptable Incorrect STOP rates), while strong competitor data could lead to more restrictive GO boundaries (low acceptable Incorrect GO rates and high acceptable Incorrect STOP rates). Also, the incorrect go rate could be lowered if the decision is gating a high investment (e.g. a large phase III programme).

Conditional operating characteristics can be obtained analytically, in some cases, or could be estimated by simulating clinical trial results over a range of fixed values of the true treatment effect, and counting the proportion of times such virtual trial results would trigger any of the possible decisions in the framework.

Figure 1 displays some graphical representations of operating characteristics for the example (a). Figure 1A shows the probabilities of GO, CONSIDER and STOP decisions for values of the true treatment effect between 0 and 4 points, for a sample size of 80 patients per arm in Phase II. Assuming there is no effect, the probability of STOP is 97% while the probabilities of CONSIDER and GO are negligible (2.6% and 0.4% respectively). When the true treatment effect is equal to the minimal value ($MV = 2$), the probability of GO is equal to 30% (pre-defined in the framework) and the probability of STOP is 41%. When the true treatment effect is equal to the target value ($TV = 3$), the probability of GO is 70.2% and the probability of STOP is 10% (pre-defined in the framework). The probability of CONSIDER decision never exceeds 30%, for all true values of the treatment effect. Overall, we can conclude that the QDM framework provides good operating characteristics here, with incorrect decision rates that are acceptable at this stage of the development. If needed, these rates could be adjusted by modifying the decision boundaries in the QDM framework or the sample size (Götte et al, 2015; Frewer et al, 2016; Quan et al, 2020). However, it may not be obvious to modify the boundaries if the decision problem is related to a business analysis (e.g. eNPV calculation). Then the choice could be to accept the risks of poor operating characteristics or increase sample size to mitigate those risks.

Figure 1B shows the probabilities of GO, CONSIDER and STOP decisions for sample sizes between 50 and 150 patients per arm, for a true treatment effect equal to the target value ($TV = 3$). For this effect, the probability of GO exceeds 70% for sample sizes greater than 80 patients per arm, with a probability of CONSIDER lower than 20% and a probability of STOP fixed at 10% (pre-defined in the framework). Such graphs are useful to support the sample size choice for achieving specific desired probabilities.

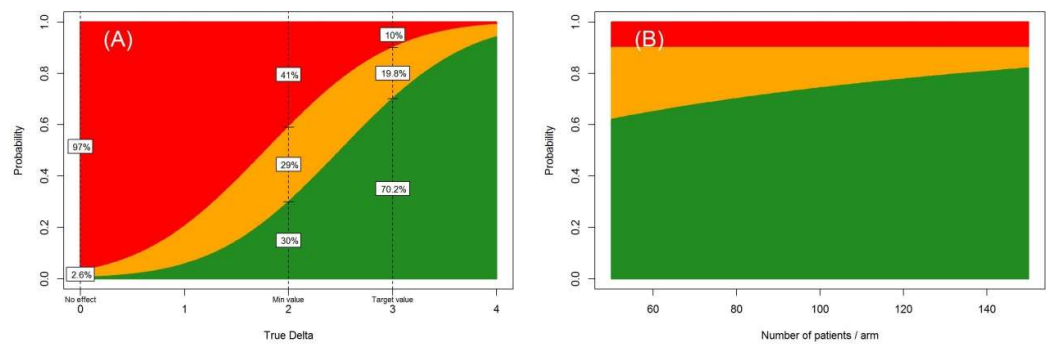

**Figure 1.** Alzheimer disease example (a). Conditional operating characteristics: probabilities of GO (green), CONSIDER (amber) and STOP (red) decisions at the end of Phase II. (A) Plotted against the true treatment effect, for N=80 patients/arm. (B) Plotted against the sample size, for a true effect=3 points (target value).

### 4.2. Unconditional operating characteristics

Although understanding the probabilities of GO, STOP and CONSIDER decisions conditional on fixed values for the treatment effect is useful, these pre-defined scenarios are not equally likely to occur. Existing data or information around the drug can be used to derive or elicit a prior distribution for the treatment effect. In a Bayesian framework, we will refer to this prior as the *design* prior, to distinguish it from the *analysis* prior $p(\delta)$ used to calculate the posterior probabilities that form the decision criteria. Then, *unconditional* operating characteristics permit to assess the average probabilities of GO, STOP and CONSIDER decisions over this design prior distribution, incorporating our uncertainty about the true treatment effect.

Still in example (a), let us assume that, based on our current knowledge, we believe the true treatment effect has a normal distribution N(3.2, 2). We can compute a weighted average of the GO/STOP/CONSIDER probabilities shown on Figure 1A with weights given by this design prior to obtain the unconditional probability of each decision outcome. Under this design prior distribution, the unconditional probabilities of GO, STOP and CONSIDER decisions at the end of the new study with 80 patients per arm are 59.4%, 31.6% and 9%, respectively.

The unconditional probabilities depend on the precision of the design prior and on how optimistic/pessimistic it is. Carroll (2013) investigated the properties of the unconditional probability of GO, or *assurance* (O'Hagan, 2005). It is expected to be close to 50% when the design prior belief is very unprecise (and thus a non-informative design prior is not helpful); it is expected to be large for an optimistic design prior, i.e. with mean above $MV$, (or, respectively, small for a pessimistic design prior); and it is expected to be close to 50% for a design prior centred on the $MV$, whatever the amount of available evidence (Saint-Hilary et al, 2018). As argued in Crisp et al (2018), no general threshold could be provided to indicate whether a level of assurance supports commitment to fund a study or not. High levels of assurance might be required in the case of a well understood disease with a highly competitive environment, while lower assurance levels could be acceptable if the drug responds to a high unmet need.

Both conditional and unconditional probabilities inform about the usefulness of a QDM framework and help optimising the study design (FDA, 2019).

Very similar operating characteristics could be produced for the QDM frameworks in example (b), prior to conduct the Phase II study. Assuming as before that the true treatment effect has a normal distribution N(3.2, 2), the unconditional probability to reach the GO criteria (statistical significance at 2.5% level one-sided and clinical relevance) in Phase III is 62.8%, and the conditional probability to reach these criteria given success in Phase II is 91.4%. These results illustrate how the Phase II "de-risks" the development of the drug. In addition, some metrics could be provided to choose the sample size in Phase III based on Phase II results, or to assess how these results affect the predictions of success in Phase III. Figure 2 displays the probability to reach the GO criteria in Phase III for 100, 200 and 300 patients per arm, against plausible values of observed difference in a Phase II study with 80 patients per arm. According to the QDM framework, a GO decision is made at the end of Phase II if this probability is greater than 70%, and a STOP decision is made if it is lower than 30%. The sample sizes in Phase II and Phase III trials could be adjusted so that the decision levels are reached for realistic observable results in Phase II.

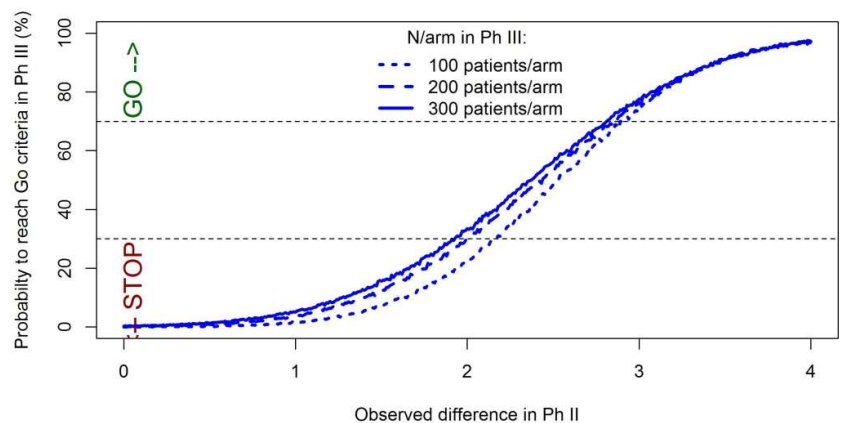

**Figure 2. Alzheimer disease example (b) for N=80 patients/arm in Phase II. Probability to reach the GO criteria (statistical significance and clinical relevance) in Phase III, versus observed difference in Phase II. A GO decision is made if it is greater than 70%. A STOP decision is made if it is lower than 30%.**

## 5. Communicating and discussing risks with study teams

This paper illustrates the important role of statistics in supporting decision-making in drug development (LaVange, 2014). However, this can only happen if statisticians succeed in communicating their value to cross-functional team members, through both strong leadership skills and effective visualization tools: '*the importance of getting the message across*' and '*the power of picture*' (Mahon, 1977).

Statisticians should pro-actively initiate and drive the discussions around quantitative decision-making. Ideally, they should take place at the planning stage (e.g. at the design phase of a study) and not retrospectively, so that the construction of the QDM framework is not influenced by the results and can be used to optimise the study design. An effective communication makes decision-makers understand how strategic quantitative thinking helps to make better decisions (Gibson, 2019), by increasing awareness of the risks in the presence of uncertainty, and by quantifying and controlling the risk of false decisions.

The first communication key is to keep the message simple, avoiding excessive details, and speak in understandable, non-technical, language. Operating characteristics are useful for this purpose, as they translate theoretical decision-making criteria into concrete study results and interpretable quantities. For example, they permit to link the posterior probability statements used to build the QDM framework to 'chances of study success' or 'observed values in the future study leading to success'.

Operating characteristics are usually better summarized and communicated through graphs. Indeed, visualization tools help people understand and interpret the frameworks, and facilitate their adoption of quantitative decision-making. They could also help in agreeing upfront with the teams on the input values (such as MV and TV) to be pre-specified in the QDM frameworks. Various types of plots could be provided, displaying the relationship between decisions and sample size, framework parameters such as the minimal effect, the target effect and the levels of acceptable risk, or observed study results.

For example, on example (a), Figure 3 maps decision criteria of the QDM framework to observed study results, in order to translate decision criteria to a more intuitive scale: the estimated value of the treatment effect. With 80 patients per arm in Phase II, assuming a certain observed standard deviation, an estimated treatment effect lower than 1.78 on the primary endpoint will lead to a STOP decision (when $\Pr(\delta < TV \mid data\ in\ Ph\ II) > 90\%$, meaning that the two-sided 80% credible interval of $\delta$ is below TV), while a value greater than 2.50 will lead to a GO decision (when $\Pr(\delta > MV \mid data\ in\ Ph\ II) > 70\%$, meaning that the two-sided 40% credible interval is just above MV). All values in between will lead to a CONSIDER decision, when more data or results on other endpoints are required to make a decision.

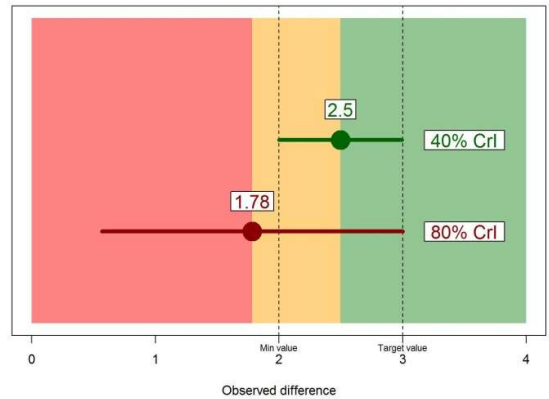

**Figure 3. Observed treatment differences leading to GO (green), CONSIDER (amber) and STOP (red) decisions at the end of Phase II, for N=80 patients/arm. CrI = Credible Interval.**

Discussions around these graphs facilitate the exchanges of statisticians with non-statisticians, in order to better capture their needs and clarify potential misinterpretations. In the end, a company adopting QDM frameworks may come to harmonised ways (templates) of presenting decision criteria to governance boards, enhancing the consistency and transparency of the decisions across the different projects.

## 6. Discussion

QDM frameworks are GO/STOP criteria specified before a study starts, or before decision milestones of a development programme or a portfolio. Such criteria are built on two key elements: the (unknown) quantity of interest $\delta$ that is most relevant for the decision at hand, and the level of uncertainty we are prepared to accept with respect of the magnitude of $\delta$. Scientific judgement coupled with existing knowledge around the therapeutic area(s), the biology of the new medicine(s), the competitor landscape, as well as other considerations (e.g. potential for return on investment), are important elements to set up a QDM framework. More complex settings involving more than one endpoint can be built following the same principles.

Once a QDM framework has been proposed, assessing its conditional operating characteristics under certain assumed values for $\delta$ will help to understand the long-run operating characteristics, i.e. the chance of a correct or an incorrect decision. This will in turn feedback to other aspects of the experiment, such as sample size. At the study level, the interplay between the elements forming the decision criteria and the sample size will inform teams about the level of evidence that can be generated and the risks associated with the choice of specific combinations of those design aspects. We note that understanding conditional operating characteristics is relevant even when the QDM criteria are based on posterior probabilities calculated with the Bayesian framework. This "calibrated" Bayes approach (Little, 2006) is the one that seems to be gaining more acceptance within the pharmaceutical industry and regulatory agencies. The FDA's draft guidance on complex innovative designs (2019) points out that calibration of Bayesian designs to frequentist operating characteristics can also help to compare a Bayesian proposal with previous studies or development programs that used frequentist inference. An alternative to the "calibrated" Bayes approach presented here could be a decision-theoretic approach based on utility functions and Bayes risks (Rosner, 2020).

Calculation of assurances is also valuable to assess the risks associated with a study or development program based on our current belief on plausible values for $\delta$. The prior distribution reflecting those beliefs can be based on previous data alone or on expert elicitation. Understanding how the assurances evolve across development phases will help to understand the value of each study in the development programme in terms of risk reduction (Temple and Robertson, 2020).

Although not common yet, the same reasoning applies at the portfolio level where pre-specified decision criteria could support the choice between development options and portfolio strategies. These QDM frameworks rely on quantitative trade-off analyses with key parameters such as the financial value of portfolio (Patel et al., 2013; Thunecke and Elze, 2014; Graham et al, 2020).

The use of QDM frameworks in pharmaceutical programs and portfolios is a valuable tool for more efficient evidence-based decision making that will ultimately reduce the current high attrition rates and optimize the investments. The pre-specification of decision criteria will help to reduce the risk of post-hoc theories to justify results, and hence the risk for wrong decisions. QDM methods are to aid, but not replace, 'human' decision making. The use of scientific knowledge and subjective judgement from the project team is crucial to understand and interpret the results. We simply advocate using it ahead of the experiments to establish decision criteria that will map outcomes to decisions. In other words, such judgement should be used to set a QDM framework, which may be automatically applied once the results are in place. In the long run, adopting QDM frameworks will reduce the chances of wrong decisions and will shorten the time between development phases. Some flexibility may be needed, for example leaving the possibility to amend the framework to take into account changes in the competition. While constructing QDM frameworks is a team effort, we encourage statisticians to drive the discussions on the grounds of their familiarity with uncertainty and its quantification.

## Acknowledgements

We thank Pierre Colin for valuable and insightful discussions.